\documentstyle[12pt]{article}
\begin{document}
\title{Finite Order BFFT Method}
\author{M. Monemzadeh$^{a}$\thanks{e-mail: monemzadeh@sepahan.iut.ac.ir} \\ A. Shirzad$^{a,b}$\thanks{shirzad@ipm.ir} \\  \\
  $^a$~{\it Department of  Physics, Isfahan University of Technology (IUT)}\\
{\it Isfahan,  Iran,} \\
 $^b$~{\it Institute for Studies in Theoretical Physics and Mathematics (IPM)}\\
{\it P. O. Box: 19395-5531, Tehran, Iran.}}
\date{}
\maketitle

\begin{abstract}
We have proposed a method in the context of BFFT approach that leads to truncation of
the infinite series regarded to constraints in the extended phase space, as well as
other physical quantities (such as Hamiltonian). This has been done for cases where
the matrix of Poisson brackets among the constraints is symplectic or constant. The
method is applied to Proca model, single self dual chiral bosons and chiral Schwinger
models as examples.
\end{abstract}
\section{Introduction}
The Dirac procedure is well-known for canonical quantization of the first class
constrained systems \cite{Dirac}. The corresponding analysis in the path integral
approach was also initiated by Faddeev for gauge theories \cite{Faddev}. To quantize a
second class constrained system in Dirac approach, it is necessary to replace Poisson
brackets by Dirac brackets. Converting Dirac brackets to quantum commutators
sometimes implies factor ordering problem and quantization of these models is not
formal. Batalin and his collaborators proposed the conversion of the second class
constraints into first class ones by defining a set of new auxiliary variables
\cite{BatFrad,BatTyu}. In this method (the BFFT method) one can find correction terms
for constraints and Hamiltonian in an iterative process, the first correction is
linear in the new variables, the second is quadratic and so on. In this way one
obtains a gauge theory and then applies the well-known mechanisms for their
quantization \cite{Faddev,FradVil,Bec,Henn}. It is important to notice that this idea
is a logical following of the original notion of St\"{u}ckelberg who converted second
class theories to first class ones by extending the configuration space with some
scaler fields (St\"{u}ckelberg scalers) \cite{Stuck}.

In this paper we show that there exist some arbitrary parameters that if suitably
chosen then the series of the correction terms of constraints and Hamiltonian do
terminate. We call this approach "the finite order BFFT method". In section 2 we
briefly review the essence of the BFFT method. Without losing the generality we assume
a system with second class constraints only. In section 3 we show that in principle it
is possible to chose the arbitrary parameters in such a way that the correction terms
terminate, provided that the matrix of Poisson brackets of constraints is either
symplectic or constant. We apply our process to the Proca model, single self dual
chiral bosons and chiral Schwinger Model in sections 4,5 and 6 respectively. Section
7 is devoted to conclusions.
\section{ Brief Review of the BFFT Formalism}
Consider a second  class constrained system described by Hamiltonian $H_0$ in phase
space with coordinates $(q^i,p_i)$ where $i=1,2,...K$. Assume the system is under the
influence of a set of second class constraints, $\Theta_\alpha \hspace
{3mm}\alpha=1,...m$, satisfying the algebra
 \begin{equation}
 \Delta_{\alpha\beta}=\{\Theta_{\alpha}, \Theta_{\beta}\}
 \label{a1}
 \end{equation}
where $ \{ , \}$ means Poisson bracket and $\Delta_{\alpha\beta}$ is an invertible
matrix. For converting a second class system into a true gauge system one can enlarge
the phase space by introducing auxiliary variables, one for each constraint. We
denote the variables by $\eta^{\alpha}$ with the following algebra:
 \begin{equation}
 \{\eta^{\alpha}, \eta^{\beta}\}= \omega^{\alpha\beta}
 \label{a2}
 \end{equation}
where $\omega^{\alpha\beta}$ is an antisymmetric matrix which we {\it assume} it to be
constant. The first class constraints in the extended phase space $(q,p)\oplus(\eta)$
are defined by
 \begin{equation}
 \tau_{\alpha}=\tau_{\alpha}(q,p,\eta)\hspace {1cm} \alpha=1,2,....,m
 \label{a3}
 \end{equation}
  with the boundary conditions
  \begin{equation}
  \tau_{\alpha}(q,p,0)=\Theta_{\alpha}(q,p).
  \label{a4}
  \end{equation}
In the abelian BFFT embedding method one demands that these
extended constraints are strongly involutive:
 \begin{equation}
 \{\tau_{\alpha},\tau_{\beta}\}=0.
 \label{a5}
 \end{equation}
The solution of the above equation can be obtained by considering $\tau_{\alpha}$ as:
 \begin{equation}
 \tau_{\alpha}=\sum_{n=0}^\infty \tau_\alpha^{(n)}
 \label{a6}
 \end{equation}
where $\tau^{(n)}_\alpha$ is of order $n$ with respect to $\eta^\alpha$'s. According
to the boundary condition (\ref{a4}) we have
 \begin{equation}
 \tau^{(0)}_\alpha=\Theta_\alpha.
 \label{a7}
 \end{equation}
Substituting Eq. (\ref{a6}) into Eq. (\ref{a5}) leads to a set of
recursive relations. Vanishing the term independent of $\eta$
gives:
\begin{equation}
 \{\tau^{(0)} _\alpha , \tau^{(0)} _\beta\}+ \{\tau^{(1)} _\alpha , \tau^{(1)}
  _\beta\}_{(\eta)}=0;
  \label{a8}
  \end{equation}
and vanishing the terms of order $n$ with respect to $\eta^\alpha$'s for $n\geq1$
gives
  \begin{equation}
  \{\tau_{[\alpha}^{(1)},\tau_{\beta ]}^{(n+1)}\}_{(\eta)}+B_{\alpha\beta}^{(n)}=0 \hspace{1cm}
  n\geq 1
  \label{a9}
  \end{equation}
where
  \begin{equation}
  B_{\alpha\beta}^{(1)}\equiv \{\tau_{[\alpha}^{(0)},
  \tau_{\beta]}^{(1)}\}
  \label{b10}
  \end{equation}
and
  \begin{equation}
  B_{\alpha\beta}^{(n)}\equiv
  \frac{1}{2}B_{[\alpha\beta]}\equiv \sum_{m=0}^n \{\tau_\alpha ^{(n-m)}, \tau_\beta
  ^{(m)}\}+\sum_{m=0}^{n-2} \{\tau_\alpha^{(n-m)}, \tau_\beta ^{(m+2)}\}_{(\eta)}\hspace
  {1cm} n\geq2.
  \label{a10}
  \end{equation}
The suffix $\eta$ in the above equations means that the Poisson brackets must be
evaluated with respect to $\eta$ variables only, otherwise they are calculated in the
basis $(q,p)$. The above equations are used iteratively to obtain the correction terms
$\tau^{(n)}$. Since $\tau^{(1)}$  is linear with respect to $\eta$ we may write
 \begin{equation}
 \tau_\alpha ^{(1)}=\chi_{\alpha\beta} (q,p)\eta^\beta.
 \label{a11}
 \end{equation}
Substituting this expression into Eq.(\ref{a8}) and using Eqs.(\ref{a1}) and
(\ref{a2}) we obtain:
 \begin{equation}
 \Delta_{\alpha\beta}+\chi_{\alpha\gamma}\omega^{\gamma\lambda}\chi_{\beta\lambda}=0.
 \label{a12}
 \end{equation}
This equation contains two unknown elements; $\chi_{\alpha\beta}$ and
$\omega^{\alpha\beta}$. One should at first assume a suitable anti-symmetric matrix
for $\omega^{\alpha\beta}$ and then solve Eq. (\ref{a12}) to determine the
coefficients $\chi_{\alpha\beta}$. Since $\Delta_{\alpha\beta}$ and
$\omega^{\alpha\beta}$ are anti-symmetric matrices, there are totally
$\frac{m(m-1)}{2}$ independent equations for $\chi_{\alpha\beta}$, while the number
of $\chi_{\alpha\beta}$'s are $m^2$. Therefore there exist an infinite number of
solutions for $\chi_{\alpha\beta}$ and we are allowed to chose any solution we wish.
Using this possibility, $\chi_{\alpha\beta}$'s can be chosen such that the process of
determining the correction terms $\tau^{(n)}$ terminate at this stage, i.e.
$\tau^{(2)}$ vanishes. We will come to this point in the next section. It can be seen
that the general solution of Eq. (\ref{a9}) is given by \cite{BanBanGh1}
 \begin{equation}
 \tau_\alpha ^{(n+1)}=-\frac{1}{n+2} \eta^\mu
 \omega_{\mu\nu}\chi^{\nu\rho}B_{\rho\alpha}^{(n)}; \hspace {1cm}n\geq1
 \label{a13}
 \end{equation}
where $\omega_{\alpha\beta}$ and $\chi^{\alpha\beta}$ are inverse to
$\omega^{\alpha\beta}$ and $\chi_{\alpha\beta}$ respectively.

To construct  the corresponding Hamiltonian $\tilde{H}(q,p,\eta)$ in the extended
phase space we demand
  \begin{equation}
  \tilde{H}(q,p,0)=H(q,p)
  \label{a14}
  \end{equation}
 and
  \begin{equation}
  \{\tau_\alpha,\tilde{H}\}=0.
  \label{a15}
  \end{equation}
Similar to $\tau_\alpha$, suppose
  \begin{equation}
  \tilde{H}=\sum_{n=0}^\infty\tilde{H}^{(n)}
  \label{a16}
  \end{equation}
where $\tilde{H}^{(n)}$ is of order $n$ with respect to $\eta^\alpha$'s and
 \begin{equation}
 \tilde{H}^{(0)}=H(q,p).
  \label{a17}
  \end{equation}
Substituting from Eqs. (\ref{a6}) and (\ref{a16}) in Eq. (\ref{a15}) gives:
   \begin{equation}
   \{\tau_\alpha^{(1)}, \tilde{H}^{(n+1)}\}_{(\eta)}+G_\alpha^{(n)}=0;\hspace {1cm}
   n\geq0
   \label{a18}
   \end{equation}
where $G_\alpha^{(n)}$ as the generators of the $\tilde{H}^{(n+1)}$ are defined as
follow
  \begin{equation}
  G_\alpha ^{(0)}\equiv\{\tau_\alpha ^{(0)}, \tilde{H}^{(0)}\}
   \label{a19}
   \end{equation}
  \begin{equation}
  G_\alpha ^{(1)}\equiv\{\tau_\alpha ^{(1)}, \tilde{H}^{(0)}\}+ \{\tau_\alpha ^{(0)},
  \tilde{H}^{(1)}\}+\{\tau_\alpha ^{(2)}, \tilde{H}^{(1)}\}_{(\eta)}
  \label{a20}
  \end{equation}
  \begin{equation}
  G_\alpha ^{(n)}\equiv \sum_{m=0}^n  \{\tau_\alpha ^{(n-m)}, \tilde{H}^{(m)}\}+
  \sum_{m=0}^{n-2}   \{ \tau_\alpha ^{(n-m)}, \tilde{H}^{(m+2)}\}_{(\eta)}+\{\tau_\alpha
  ^{(n+1)}, \tilde{H}^{(1)}\}_{(\eta)}; \hspace {0.5 cm} n\geq2.
  \label{a21}
  \end{equation}
It can be shown that the general expression for $\tilde{H}^{(n)}$ is
 \begin{equation}
 \tilde{H}^{(n+1)}=-
 \frac{1}{n+1}\eta^\alpha\omega_{\alpha\beta}\chi^{\beta\nu}G^{(n)}_\nu.
 \label{a22}
 \end{equation}
Similarly for every function $F(q,p)$ in the phase space one can
write
 \begin{equation}
 \tilde{F}(q,p,\eta)=\sum_{n=0}^\infty\tilde{F}^{(n)},
  \label{a23}
  \end{equation}
where $\tilde{F}^{(n)}$ is of order $n$ with respect to $\eta^\alpha$'s and
 \begin{equation}
 \tilde{F}^{(n+1)}
 =-\frac{1}{n+1}\eta^\alpha\omega_{\alpha\beta}\chi^{\beta\nu}\rho_\nu ^{(n)}.
 \label{a24}
 \end{equation}
In this relation $\rho_\nu ^{(n)}$ can be derived similar to $G_\nu ^{(n)}$ in Eqs.
(\ref{a19}-\ref{a21}) by replacing $H$ with $F$.

This completes the BFFT construction of the first class system which is strongly
involutive. As can be seen the correction terms of $\tau_\alpha ^{(n)}$ and
$\tilde{H}^{(n)}$ are derived iteratively from Eqs.(\ref{a13}) and (\ref{a22}).
Generally, there is no guarantee that the series terminate at some definite order.
However, the series will terminate if $B_{\alpha\beta}^{(N)}$ and $G_\alpha^{(N)}$
vanish for a certain order $n=N$.
\section{Finite Order Method}
In this section we want to solve the iterative equations for $\tau_\alpha^{(n)}$ and
$\tilde{H}^(n)$ in such a way that the corresponding series terminate as soon as
possible. We remember that $\omega^{\alpha\beta}$ can be chosen arbitrarily. On the
other hand Eq. (\ref{a12}) for $\chi_{\alpha\beta}$'s is not so much restrictive. We
use these possibilities to find a systematic method to truncate infinite series
encountered in BFFT method. However, the problem seems difficult for a general second
class system. In the following we solve it for two special cases, i.e. where the
matrix $\Delta_{\alpha\beta}$ given in (\ref{a1}) is symplectic or constant.

\vspace{5mm}\textbf{A-} Suppose $\Delta_{ij}=J_{ij}$, where $J$ is the symplectic
matrix: $$J= \left(
 \begin{array}{c|c}
 \textbf{0}&-\textbf{1} \\ \hline +\textbf{1}&\textbf{0}
 \end{array}\right). $$ In principle it has been shown that one can usually redefine the second
class constraints as pairs of coordinates and momenta with the symplectic algebra
\cite{Loran2,LoranShir1}. The algebra of the new variables $\eta^\alpha$ and unknown
coefficients $\chi_{\alpha\beta}$ can be chosen as
 \begin{equation}
 \begin{array}{l}
 \omega^{\alpha\beta}=\{\eta^\alpha,\eta^\beta\}=\tilde{J}_{\alpha\beta}=-J_{\alpha\beta}\\
 \chi_{\alpha\beta}=J_{\alpha\beta}
 \end{array}
 \label{a25}
 \end{equation}
It is easy to check that $\omega$ and $\chi$ in Eq. (\ref{a25}) satisfy the basic
equation Eq. (\ref{a12}) for $\Delta=J$. So the first correction term of the
constraints is
 \begin{equation}
  \tau_\alpha ^{(1)} =\chi_{\alpha\beta} \eta^\beta =J_{\alpha\beta}\eta^\beta.
 \label{a26}
 \end{equation}
Since $\tau_\alpha ^{(1)}$ is only a function of $\eta$, it can be seen in a
straightforward way that $B_{\alpha\beta}^{(n)}$ vanish for $n\geq1$. As a result
$\tau_\alpha$ series terminate at this step. The new set of constraints are found to
be
 \begin{equation}
  \tau_\alpha (q,p,\eta) = \tau_\alpha (q,p)^{(0)} + J_{\alpha\beta}\eta^\beta.
 \label{a27}
 \end{equation}
One can directly check that $\tau_\alpha$'s are strongly involutive. To complete our
procedure we should also construct the extended Hamiltonian. Inserting (\ref{a25})
into (\ref{a22}), the correction terms of Hamiltonian are deduced as
 \begin{equation}
 \tilde{H}^{(n+1)}=-\frac{1}{n+1} \eta^\alpha G_\alpha^{(n)}
 \label{aa27}
 \end{equation}
It is necessary to evaluate the $G_\alpha^{(n)}$ as the generators of $\tilde{H}$;
i.e. $\tilde{H}^{(n+1)}\sim G_\alpha ^{(n)}$. For the the zeroth order we have
 \begin{equation}
 G_\mu ^{(0)}=\{\tau_\mu^{(0)} , H_0\}.
 \label{a28}
 \end{equation}
 The next correction term for $\tilde{H}$ is
 \begin{equation}
 \tilde{H}^{(1)} = -\eta^\mu G_\mu ^{(0)}.
 \label{a29}
 \end{equation}
This should be inserted into Eq. (\ref{a20}) to find
 \begin{equation}
 G_\alpha ^{(1)}=-\eta^\mu C_{\alpha\mu}
 \label{a30}
 \end{equation}
where
 \begin{equation}
 C_{\alpha\mu}=\{\tau_\alpha ^{(0)} , \{\tau_\mu ^{(0)} , H_0\}\}.
 \label{a31}
 \end{equation}
Similarly $\tilde{H}^{(2)}$can be derived from Eq. (\ref{a22}) as
  \begin{equation}
  \tilde{H}^{(2)} = \frac{1}{2} \eta^\mu \eta^\nu C_{\mu\nu}.
  \label{a32}
  \end{equation}
This process continue until $G_\alpha ^{(n)}$ become a function of $\eta$'s only. If
$H_0$ is at most quadratic with respect to phase space coordinates, it would be clear
that $C_{\alpha\mu}$ in Eq. (\ref{a31}) is constant and $G_\alpha ^{(2)}=0$; and
consequently $\tilde{H}^{(3)}=0$. In this case one can finally write
 \begin{equation}
 \tilde{H}=H_0-\eta^\mu C_\mu ^{(0)}+ \frac{1}{2}\eta^\mu \eta^\nu C_{\mu\nu}.
 \label{a33}
 \end{equation}
 In a more general case, when $H_0$ is a function of order $N$ with respect to coordinates
 $(q,p)$ and the constraints are linear with respect to coordinates and momenta, the series of $\tilde{H}$
 will be finished at $N$th step; i.e.
 \begin{equation}
 \tilde{H}=H_0+\tilde{H}^{(1)}+...+\tilde{H}^{(N)}
 \label{a34}
 \end{equation}
Eqs. (\ref{a27}) and (\ref{a34}) represent a finite order gauge theory in abelian
BFFT approach. In this way we can convert every second class constraint system to a
{\it rank zero gauge theory}, in which the structure functions $C_{\alpha\beta}
^\gamma$ and $V_\alpha ^\beta$ defined in
 \begin{equation}
 \begin{array}{l}
 \{\tau_\alpha , \tau_\beta\} = C_{\alpha\beta} ^\gamma \tau_\gamma \\ \{\tau_\alpha , \tilde{H} \} = V_\alpha ^\beta
 \tau_\beta
 \end{array}
 \label{a35}
 \end{equation}
vanish in the extended phase space \cite{Henn}.

Assuming again that $\Delta$ is the symplectic matrix, one can also select
$\omega=\Delta^T=-J$. Then the basic Eq. (\ref{a12}) implies that
 \begin{equation}
 J=\chi^T J \chi.
 \label{C1}
 \end{equation}
As stated in Eq.(\ref{a25}), $\chi=J$ satisfy the above equation. On the other hand,
as is well-known \cite{Gold}, a canonical transformation from the set $(q,p)$ to
$(Q,P)$ is represented by
 \begin{equation}
 J=M^T J M
 \label{CC1}
 \end{equation}
where
 \begin{equation}
 M=\frac{\partial(q,p)}{\partial(Q,P)}.
 \label{CC2}
 \end{equation}
Comparing Eq. (\ref{CC1}) with Eq. (\ref{C1}) shows that any canonical transformation
in phase space of $(q,p)$ can introduce a solution to the basic equation (\ref{a12}).
In this way a large class of solutions are obtained, among them those with constant
elements for $M$ give truncated series for constraints.

\vspace{5mm} \textbf{B-} In most physical examples of second class systems the
$\Delta$-matrix in (\ref{a1}) emerge as a matrix with constant elements. In this case
we can choose
 \begin{equation}
 \omega=\Delta^T=-\Delta.
 \label{C2}
 \end{equation}
So the basic Eq. (\ref{a12}) can be written as
 \begin{equation}
 \Delta-\chi^T \Delta\chi=0.
 \label{C3}
 \end{equation}
It is easy to see that $\chi=1$ satisfies the above equation. Then the new set of
constraints are of the form
 \begin{equation}
 \tau_\alpha=\tau_\alpha ^{(0)}+\eta^\alpha.
 \label{C4}
 \end{equation}
The correction terms of the Hamiltonian can be derived as\footnote{Notice that the
indices $\alpha, \beta, ...$ have not tensorial mining. i.e. there is no metric to
rase up or lower down the indices. Therefore the reader should not be worried
 about up-down indices on matrix $\Delta^{-1}$ in Eq. (\ref{C5}), etc.}
 \begin{equation}
 \tilde{H}^{(n+1)}=\frac{1}{n+1}\eta^\alpha(\Delta^{-1})_\alpha ^\beta G_\beta
 ^{(n)}
 \label{C5}
 \end{equation}
where $G_\alpha ^{(n)}$ are defined in Eqs. (\ref{a19}-\ref{a21}). For a Hamiltonian
which is a polynomial of order $N$ with respect to the original phase space
coordinates $(q,p)$, the generators $G_\alpha ^{(N)}$ will be only a function of
auxiliary variables. Therefore the $\tilde{H}$ series will terminate at $N$th step and
the constraints (\ref{C4}) and $\tilde{H}$ with correction terms (\ref{C5}) represent
a rank zero gauge theory.

The significance of the above method can be better seen in the context of the chain by
chain method introduced recently in \cite{LoranShir1}. Suppose we have only one chain
of second class constraints with the recursion formula:
 \begin{equation}
 \Theta_{n+1}=\{\Theta_n , H_0\}.
 \label{g1}
 \end{equation}
Suppose $\Delta$ is a matrix with constant elements and we choose our arbitrary
parameters $\omega$ and $\chi$ in such a way that the new set of constraint are given
by Eq.(\ref{C4}). It is clear from (\ref{g1}) and (\ref{a19}) that
 \begin{equation}
 \begin{array}{l}
 G_\alpha ^{(0)}=\Theta_{\alpha+1} \hspace{1.5cm} \alpha=1,2,...,m-1 \\ G_m ^{(0)}
 =\{\Theta_m , H_0\}.
 \end{array}
 \label{g2}
 \end{equation}
So the first correction term of $\tilde{H}$ is
 \begin{equation}
 \tilde{H}^{(1)}=\sum_{\beta=1}^{m-1} \eta^\alpha (\Delta^{-1})_\alpha ^\beta
 \Theta_{\beta+1} + \eta^\alpha(\Delta^{-1})_\alpha ^m \{\Theta_m , H_0\}.
 \label{g3}
 \end{equation}
It can be seen that
 \begin{equation}
 \tilde{H}^{(2)}=\sum_{\beta=1}^{m-1} \frac{1}{2} \eta^\mu \eta^\alpha (\Delta^{-1})_\mu
 ^\nu (\Delta^{-1})_\alpha ^\beta
 \Delta_{\nu \beta+1} + \frac{1}{2}\eta^\mu \eta^\alpha (\Delta^{-1})_\mu ^\nu(\Delta^{-1})_\alpha ^m \{\Theta_\nu , \{\Theta_m , H_0\}\}.
 \label{g4}
 \end{equation}
As we know from Eq.(\ref{a22}); $\tilde{H}^{(n+1)}\sim G_\alpha ^{(n)}$ and
 \begin{equation}
 G_\alpha ^{(n)} \sim \{\Theta_\alpha , \{\Theta_{\alpha_1} , \{\Theta_{\alpha_2} ,
 ...\{\Theta_{\alpha_n} , H_0\}\}...\}.
 \label{g5}
 \end{equation}
If $H_0$ is a polynomial of finite order $N$ with respect to the phase space
coordinates, then Eq. (\ref{g5}) shows that its correction terms do terminate at most
after $N$ steps.

Now we apply the above procedures to some definite models.
\section{The Proca Model}
As the first example we consider the Proca model, whose dynamics is described by the
Lagrangian density
 \begin{equation}
 {\cal{L}}= -\frac{1}{4} F^{\mu\nu} F_{\mu\nu} + \frac{1}{2} A^\mu
 A_\mu
 \label{a36}
 \end{equation}
where
 \begin{equation}
  F^{\mu\nu}=\partial^\mu A^\nu - \partial^\nu A^\mu.
 \label{a37}
 \end{equation}
It is well-known that the second term in Eq. (\ref{a36}) breaks the gauge symmetry of
the usual Maxwell's theory given by the first term. The canonical momenta are defined
as
 \begin{equation}
 \pi^\mu(x) = \frac{\partial {\cal{L}}}{\partial
 \dot{A_\mu}}=-F^{0\mu}(x).
 \label{a38}
 \end{equation}
From Eqs. (\ref{a37}) and (\ref{a38}) there is only one primary constraint field
  \begin{equation}
  \Theta_1(x) \equiv \pi^0(x)\approx 0
  \label{a39}
  \end{equation}
 where the symbol $\approx $ means weak equality. The canonical Hamiltonian is
  \begin{equation}
  H_c=\int\left[\frac{1}{2}\pi_i ^2 + \frac{1}{4}F_{ij} ^2+\frac{1}{2}(A_i^2 - A_0 ^2)-A_0
  \partial_i \pi^i\right] d\textbf{x}.
  \label{a40}
  \end{equation}
The total Hamiltonian is defined as
 \begin{equation}
 H_T=H_c+\int d\textbf{x} \lambda(x)  \Theta_1(x)
  \label{a41}
  \end{equation}
where $\lambda(x)$ is the Lagrange multiplier field.
 Following the algorithm of Dirac, we find that the consistency in time of the primary constraint (i.e. $\dot{\Theta_1}=\{\Theta_1 ,
H_C\}=0$) leads to the secondary constraint field
 \begin{equation}
  \Theta_2(x) \equiv \partial_i \pi^i +A_0 \approx0.
  \label{a42}
  \end{equation}
The consistency condition of $\Theta_2$ just determines the Lagrange multiplier
$\lambda(x)$. The algebra of the second class constraints in Eqs. (\ref{a39}) and
(\ref{a42}) satisfy the basic condition\footnote{Since the constraints are space-time
fields, a three dimensional Dirac $\delta$-function should be understood in Poisson
bracket of constraints. More precisely we have
 $$\{\Theta_i(\textbf{x},t) , \Theta_j(\textbf{y},t)\}=\delta(\textbf{x}-\textbf{y}) J_{ij} \hspace{1.5cm} i,j=1,2.$$
However, we omit the $\delta$-functions when not needed.} $\Delta_{ij}=J_{ij}$. For
simplicity in our calculation we apply the following canonical transformation
  \begin{equation}
  \begin{array}{l}
  A=\partial_i \pi^i +A_0 \\ \pi=\pi^0 \\  A' _i =A_i + \partial_i \pi^0 \\ \pi'^i
  = \pi^i.
  \end{array}
  \label{a43}
  \end{equation}
The new set of constraints and Hamiltonian are found to be
  \begin{equation}
   \Theta'_1(x)\equiv A(x)=0 \hspace{2.5cm} \Theta'_2(x)\equiv\pi(x)=0
  \label{a44}
  \end{equation}
  \begin{equation}
  H'_c=\frac{1}{2}\pi'^{i} \pi'
  _{i}+\frac{1}{4}\tilde{F}^{ij} \tilde{F}_{ij}+\frac{1}{2}\left[(\partial_i
  \pi'^i)^2+(\partial_i \pi)^2 -A^2\right]-A'_i \partial_i\pi
  \label{b44}
  \end{equation}
where
 \begin{equation}
 \tilde{F}^{ij}=\partial^i A'^j - \partial^j A'^i.
 \label{a45}
 \end{equation}
 In order to convert the above gauge non-invariant theory to a first class one, we
 make use of two new auxiliary fields $\eta^1$ and $\eta^2$. According to Eq. (\ref{a25}) we
 choose
 \begin{equation}
 \begin{array}{l}
 \omega^{\alpha\beta}=\{\eta^\alpha , \eta^\beta\}=-J^{\alpha\beta} \\ \chi_{\alpha\beta}=J_{\alpha\beta}.
 \end{array}
 \label{a46}
 \end{equation}
The first class constraints are deduced from Eq. (\ref{a27}) as
 \begin{equation}
 \tau_1 \equiv A+\eta^2 \hspace{2.5cm} \tau_2 \equiv \pi - \eta^1.
 \label{a47}
 \end{equation}
The generators in the first correction term of $H'_c$ are
 \begin{equation}
 \begin{array}{l}
 G_1^{(0)}= \partial_i A_i'(x)-\partial_i\partial_i\pi(x) \\ G_2
 ^{(0)}= A(x)
 \end{array}
 \label{a48}
 \end{equation}
and from Eq. (\ref{a29}) one finds that
 \begin{equation}
 \tilde{H}'^{(1)}=\eta^1(\partial_{i}\partial_i \pi -\partial_iA'_{i})-\eta^2 A.
 \label{a49}
 \end{equation}
Explicit calculations from Eq. (\ref{a32}) yield the last correction term as
 \begin{equation}
 \tilde{H}'^{(2)}=-\frac{1}{2}(\eta^1\partial_i\partial_i\eta^1+\eta^2\eta^2).
 \label{a50}
 \end{equation}
So the embedded Hamiltonian is
 \begin{equation}
 \tilde{H}'_C =H' _C + \tilde{H}'^{(1)} + \tilde{H}'^{(2)}.
 \label{a51}
 \end{equation}
One can easily check that Eqs. (\ref{a47}) and (\ref{a51}) represent an abelian gauge
theory.
\section{Gauge-Invariant Single Self Dual Chiral Bosons}
The gauge non-invariant Srivastava model for single self dual Chiral bosons in
$(1+1)$ dimensions is described by the Lagrangian density \cite{Srivastava}:
 \begin{equation}
 {\cal{L}}^N =\frac{1}{2}\dot{\phi}^2-\frac{1}{2}\phi'^2+\lambda(\dot{\phi}-\phi')
 \label{a52}
 \end{equation}
 where $\dot{\phi}\equiv \partial_0\phi$ and $\phi'\equiv\partial_1\phi$.

In this section we use the Lorentz metric $g^{\mu\nu}=$diag (+1, -1). The canonical
momenta can be derived as:
 \begin{equation}
 \pi=\dot{\phi}+\lambda  \hspace{1.5cm} P_\lambda=0
 \label{a53}
 \end{equation}
where $\pi$ and $P_\lambda$ are the momenta conjugate to the fields $\phi$ and
$\lambda$ respectively. There is one primary constraint $(\Theta_1 \equiv
P_\lambda\approx0)$. The canonical Hamiltonian density corresponding to ${\cal{L}}^N$
is
 \begin{equation}
 {\cal{H}}^N_C=\frac{1}{2}(\pi-\lambda)^2+\frac{1}{2}\phi'^2+\lambda\phi'.
 \label{a54}
 \end{equation}
Consistency condition of the primary constraint leads to a
secondary constraint
 \begin{equation}
 \Theta_2\equiv \pi - \phi' - \lambda \approx0.
 \label{aa54}
 \end{equation}
Since $\left[\Theta_1 , \Theta_2\right]\neq0$ the constraint chain finishes at this
step. We have two second class constraints satisfying the symplectic algebra which
represent a gauge non-invariant model. This model was considered in St\"{u}ckelberg
method with enlarging the Hilbert space
 of the theory and introducing a full quantum field $\theta$, called Wess- Zumino field \cite{WessZum}, to obtain the modified Lagrangian density as:
 \begin{equation}
 {\cal{L}}^I={\cal{L}}^N+{\cal{L}}^{WZ}; \hspace{1cm}
 {\cal{L}}^{WZ}=-\frac{1}{2}(\dot{\theta}+\theta'^2)+\theta'(\phi'+\dot{\theta})-\dot{\theta}(\phi'+\lambda)+\lambda\theta'.
 \label{a55}
 \end{equation}
In this section we concentrate on this model in BFFT method and
introduce $\eta^1$ and $\eta^2$ as auxiliary fields with the
algebra
 \begin{equation}
 \omega^{\alpha\beta}=\{\eta^\alpha,\eta^\beta\}=-J^{\alpha\beta}.
 \label{a56}
 \end{equation}
According to the procedure defined before and Eq.(\ref{a27}) the new abelian first
class constraints are:
 \begin{equation}
 \begin{array}{l}
 \tau_1\equiv P_\lambda+\eta^2 \\ \tau_2\equiv
 \pi-\phi'-\lambda-\eta^1.
 \end{array}
 \label{a57}
 \end{equation}
The embedded Hamiltonian density in the extended phase space with
the mention to (\ref{a28}-\ref{a32}) are derived as
 \begin{equation}
 \tilde{{\cal{H}}}={\cal{H}}_C^N+\tilde{{\cal{H}}}^{(1)}+\tilde{{\cal{H}}}^{(2)}
 \label{a58}
 \end{equation}
where
 \begin{equation}
 \begin{array}{l}
 \tilde{{\cal{H}}}^{(1)}= - \eta^1 ( \pi - \phi'- \lambda ) - \eta^{2} ( \phi'' + 2 \lambda' - \pi' ) \\
 \tilde{{\cal{H}}}^{(2)}= \frac{1}{2} \eta^1 \eta^1 + \eta^1
 \eta^{2'}-\eta^{1'}\eta^2-\eta^2\eta^{2''}.
 \end{array}
 \label{a59}
 \end{equation}
First class constraints (\ref{a57}) and Hamitonian (\ref{a58})
represent a rank zero gauge theory.
\section{Gauge Invariant Chiral Schwinger Models}
In this section we use our formalism in a theory in which the $\Delta$-matrix has
constant elements. The gauge non-invariant bosonized chiral Schwinger model
\cite{JacRaj,MitraRaj}, in $(1+1)$ dimensions with regularization parameter $a=1$ is
described by the Lagrangian density:
 \begin{equation}
{\cal{L}}^N = \frac{1}{2} \partial_\mu \phi \partial^\mu
  \phi + (g^{\mu\nu} - \varepsilon^{\mu\nu}) \partial_\mu\phi
  A_\nu - \frac{1}{4} F_{\mu\nu} F^{\mu\nu} + \frac{1}{2} A_\mu
  A^\mu
 \label{a60}
 \end{equation}
in which $\phi$ is a scalar field and $A_\mu$ is a vector field. There appear four
second class constraints \cite{KulKul}:
  \begin{equation}
  \begin{array}{l}
  \Theta_1 \equiv \pi_0 \approx 0 \hspace{1.5cm} \Theta_2\equiv
  E' + \phi' + \pi + A_1 \approx 0 \\ \Theta_3 \equiv E\approx 0
  \hspace{1.5cm} \Theta_4 \equiv -\pi - \phi' - 2A_1 + A_0 \approx
  0
  \end{array}
  \label{a61}
  \end{equation}
where $\pi$, $\pi_0$ and $E$ are momenta conjugate to $\phi$, $A_0$, and $A_1$
respectively. The canonical Hamiltonian density corrsponding to Eq. (\ref{a60}) is
  \begin{equation}
  {\cal{H}}^N_C = \frac{1}{2} \pi^2 + \frac{1}{2}\phi'^2 +
  \frac{1}{2}E^2 + EA'_0 + (\pi + \phi' + A_1)(A_1 - A_0).
  \label{a62}
  \end{equation}
It is clear that Eq. (\ref{a61}) represent a second class constrained system with the
algebra
 \begin{equation}
 \{\Theta_i(\textbf{x},t) , \Theta_j(\textbf{y},t)\}=\Delta_{ij}\delta(\textbf{x}-\textbf{y})
 \label{aa62}
 \end{equation}
where
 \begin{equation}\Delta= \left(
 \begin{array}{cccc}
 0&0&0&-1 \\0&0&1&0 \\0&-1&0&2 \\ 1&0&-2&0
 \end{array}\right).
 \label{a63}
 \end{equation}
As before we can omit the $\delta$-function and discuss about the discrete part
$\Delta$. For construction a first class theory, it is necessary to define four
auxiliary fields $\eta^\alpha(x)$ where $\alpha=1,2,3,4$. In agreement with Eq.
(\ref{C2}) we chose them such that
 \begin{equation}\omega^{\alpha\beta}= -\Delta= \left(
 \begin{array}{cccc}
 0&0&0&1 \\0&0&-1&0 \\0&1&0&-2 \\ -1&0&2&0
 \end{array}\right).
 \label{a64}
 \end{equation}
Remembering that the trivial choice $\chi=1$ satisfy (\ref{a12}) and  according to
(\ref{C4}), the new set of constraint are found to be:
  \begin{equation}
  \begin{array}{l}
  \tau_1 \equiv \pi_0 + \eta^1 \approx 0 \\ \tau_2\equiv
  E' + \phi' + \pi + A_1 +\eta^2 \approx 0 \\ \tau_3 \equiv E + \eta^3 \approx 0 \\
  \tau_4 \equiv -\pi - \phi' - 2A_1 + A_0 + \eta^4 \approx
  0
  \end{array}
  \label{a65}
  \end{equation}
From Eqs. (\ref{g3}) and (\ref{g4}) the correction terms of the embedded Hamiltonian
density are derived. As a result
  \begin{equation}
  {\tilde{\cal{H}}}={\cal{H}^N_C} + {\tilde{\cal{H}}^{(1)}} +{\tilde{\cal{H}}^{(2)}}
  \label{a66}
  \end{equation}
where
  \begin{equation}
  {\tilde{\cal{H}}^{(1)}}= \eta^1 \left[2\Theta_3 - \phi'' - \pi' -
  2(A'_1+E)\right]-\eta^2 (2\Theta_2 +\Theta_4) + \eta^3 \Theta_3 -
  \eta^4 \Theta_2
  \label{a67}
  \end{equation}
and
  \begin{equation}
  {\tilde{\cal{H}}^{(2)}}= 2\eta^{1} \eta^{1} - \eta^{1} \eta^{1''} -
  \eta^{1}
  \eta^{2'} - \eta^{2} \eta^{2} + \eta^{2} \eta^{1'} - \eta^{2} \eta^{4} +
  \frac{1}{2}\eta^{3} \eta^{3}.
  \label{a68}
  \end{equation}
It can be checked that Eqs. (\ref{a65}) and (\ref{a66}) represent a rank zero gauge
invariant theory in the extended phase space.
\section{Conclusion}
As discussed in the previous sections, BFFT approach is a method for converting a
second class constrained system to a first class one which can be quantized according
to the usual quantization methods of first class systems; for instance canonical
quantization or path integral approach. In this method the series of correction terms
for constraints and every function in phase space, in principle have infinite terms.
In the master equation of BFFT method, Eq.(\ref{a12}), there exist arbitrariness for
some basic parameters. It is possible to make truncated series for functions of the
phase space provided that we chose these parameters in a convenient way. This is done
for some special models in Refs. \cite{BanBanGh1,BanBanGh2}, but a systematic method
has not been proposed for the general case. However, it seems difficult to truncate
series in BFFT method for an arbitrary second class system; i.e. a system in which
the elements of $\Delta$-matrix are functions of phase space. We solved this problem
when $\Delta$- matrix is in symplectic form or its elements are constants. These
cases for $\Delta$-matrix do not lose the generality of the problem. In fact it has
been shown that one can convert every second class constrained system to a symplectic
system \cite{Loran2,LoranShir1}. On the other hand, in most covariant physical models
the $\Delta$-matrix has constant elements as we showed for some of them in sections
4-6. The method can be applied to several second class systems in the similar way.
\section*{Acknowledgment}
The authors thanks F. Loran for useful discussion and comments.

\end{document}